\documentclass[twocolumn,aps]{revtex4}
\usepackage{graphics}

\begin{document}

\title{Improvement of experimental data via consistency conditions}

\author{G. Kontrym-Sznajd}
\affiliation{W.Trzebiatowski Institute of Low Temperature and Structure
Research, Polish Academy of Sciences, P.O.Box 937, 50-950 Wroc{\l}aw 2, Poland}

\begin{abstract}
\vspace{0.5cm}
   Interdependencies between experimental spectra, representing line or plane
   projections of electronic densities, are derived from their consistency
   and  symmetry conditions. Some additional relations for plane
   projections are obtained by treating them as line projections of  line
   projections.

   The knowledge of these dependencies can be utilised both for an {\em improvement}
   of experimental data and for a verification of various techniques used for
   correcting e.g. two-dimensional (or one-dimensional) angular correlation of
   annihilation radiation spectra and Compton  scattering profiles.

   \vspace{0.5cm}
   {\bf published in:} Appl. Phys. A {\bf 70}, 97-100 (2000)
\end{abstract}


\maketitle

\section{Introduction}

 All spectra, representing projections of the same density, must be interdependent.
 This so-called consistency condition (\textit{CC}) has been considered
 for reconstructing densities $\rho({\bf p})$  from their line projections [1-3].
 In the case of angular correlation of annihilation radiation
 ($ACAR$) spectra or Compton profiles ({\em CP}) measurements this condition
 is satisfied if spectra are measured up to such a momentum $p_{max}$ above
 which $\rho({\bf p})$ is isotropic and all projections have the same values.

The {\em CC} is  automatically imposed on the experimental data via the reconstruction
of $\rho({\bf p})$. However, it can be utilised for checking (before the reconstruction)
if data were measured and next corrected (to remove various experimental imperfections)
properly. Moreover, this condition can be also profitable
for an {\em improvement} of such experimental data for  which $\rho({\bf p})$
is not reconstructed. For this purpose one should estimate interdependences
between projections what is a subject of this paper.

In the next Section we discuss the consistency condition in its
general form, i.e. for the Radon transform in {\em N}-dimensional space.
Next, some relations between line (Sec.~II.A) and plane projections (Sec.~II.B)
are derived  from both the {\em CC} and a symmetry of $\rho({\bf p})$
(electronic densisty in the momentum space) for various crystallographical structures.
A fulfilment of these  relations, proved for various model  and  experimental
spectra, is discussed in Sec.~III.

\section{Theory}
The Radon transform [4] represents integrals of $\rho({\bf p})$ (defined in
$N$-dimensional space, ${\cal R}^{N}$) over $(N-1)$-dimensional hyperplanes:
\begin{equation}
\label{e1}
\hat{R}\cdot\rho({\bf p}) = g(t,\mbox{\boldmath $\zeta$})
                          = \int_{-\infty}^{\infty}\rho({\bf p})
\delta(t-\mbox{\boldmath $\zeta$}\cdot{\bf p})d{\bf p},
\end{equation}
where $\mbox{\boldmath $\zeta$}$ is a unit vector in ${\cal R}^{N}$ along $t$ and
$t$ is a perpendicular distance of the hyperplane from the origin of the
coordinate system. The equation $t=\mbox{\boldmath $\zeta$}\cdot{\bf p}$
defines the hyperplane. In the same coordinate system the vector ${\bf p}$
is described by $p=\vert{\bf p}\vert$ and $\mbox{\boldmath $\omega$}$
(e.g. Dean [5]).

Both functions, $g$ and $\rho$, can be expanded into spherical harmonics of
degree $l$:
\begin{equation}\label{e2}
g(t,\mbox{\boldmath $\zeta$})=\sum_{l\nu}g_{l\nu}(t)S_{l\nu}
(\mbox{\boldmath $\zeta$}),
\end{equation}
\begin{equation}\label{e3}
\rho(p,\mbox{\boldmath $\omega$})=\sum_{l\nu}\rho_{l\nu}(p)S_{l\nu}
(\mbox{\boldmath $\omega$}),
\end{equation}
where index $\nu$ distinguishes harmonics of the same order $l$. In order to make our
formuale clearer, henceforth index $\nu$ will be omitted, keeping in mind that for the same $l$ a
few harmonics can be used. According to Dean [5],

\begin{equation}
\label{e4}
\rho_{l}(p)=\frac{c}{p}\int_{p}^{\infty}g_{l}^{(2\mu+1)}(t)C_{l}^{\mu}
(t/p)[(t/p)^{2}-1]^{\mu-1/2}dt,
\end{equation}
where $c=(-1)^{2\mu+1}\Gamma(l+1)\Gamma(\mu)/(2\pi^{\mu+1}\Gamma(l+2\mu))$.
Here $g^{(n)}$ denotes the $n^{th}$ derivative of $g$ ; $C^{\mu}_{l}$
are Gegenbauer polynomials and $\mu=N/2-1$. The equation (4) can be solved
analitically if either [6]
\begin{equation}\label{e5}
g_{l}(t)=\sum^{\infty}_{k=0}a_{lk}(1-t^{2})^{\lambda-1/2}
C^{\lambda}_{l+2k}(t),
\end{equation}
with $\lambda > N/2-1$, or [7]
\begin{equation}\label{e6}
g_{l}(t)=e^{-t^{2}}\sum^{\infty}_{k=0}b_{lk}H_{l+2k}(t),
\end{equation}
where $H_{m}$ are the Hermite polynomials. Because in $\rho_{l}(p)$ all terms $t^{n}$
 with  $0\le n < l$ are equal to zero,  the lowest term in $g_{l}$ is of
 order $l$. This property, called {\em CC}, follows from the fact that
 functions $g_{l}$ (Eq. (2)) are a linear combination of $g(t,\mbox{\boldmath $\zeta$})$
 which in turn are interdependent being projections of the same density.
 Generally, one can write
\begin{equation}\label{e7}
g_{l}(t)=\sum^{\infty}_{k=0}c_{lk}W_{l+2k}(t),
\end{equation}
where $W_{n}$ denotes an arbitrary orthogonal polynomial.  Below, considering the cases
$N=2$ (line projections) and $N=3$ (plane projections),  we show that due
to this property of $g_{l}(t)$ (its minimal number of zeros) we can get some
interdependences between projections $g(t,\mbox{\boldmath $\zeta$})$. For that
it is necessary to expand them into a series of the same (as in Eq.~(7))
polynomials:
\begin{equation}\label{e8}
g(t,\mbox{\boldmath $\zeta$})=\sum^{\infty}_{k=0}c_{k}(\zeta)W_{2k}(t).
\end{equation}
By combining Eqs.~(2), (7) and (8) we can obtain some relations between $c_{k}(\zeta)$
as a function of $\mbox{\boldmath $\zeta$}$. In the paper, for calculating the
coefficients $c_{k}$,
the  Chebyshev polynomials and the Gaussian quadrature formulae were applied.

\subsection{Line projections}

In the case of reconstructing $\rho({\bf p})$  from its line projections
$g(x,y)$ (measured in the apparatus systems $(x,y,z)$), the reconstruction of
the three--dimensional ($3D$) density can be reduced to a set of reconstructions
of $2D$ densities (the Radon transform for $N=2$) performed, independently,
on succeeding planes $y=const.$, parallel each other. In order to describe all
projections in the same coordinate
system (in which the symmetry of both $\rho({\bf p})$ and measured spectra is
defined)  the function $g$ can be characterized in the polar system where
$g(x,y)\equiv g(t,\alpha)$.  Here $t$ and $\alpha$ denote the
distance of the  integration line from the origin of the coordinate
system and its angle with  respect to a chosen axis, respectively.

For the planes $y=const.$, perpendicular to the axis of the crystal rotation
of the order $\vert G\vert$, Eq. (2)  reduces to the $cosine$ series [8]:
\begin{equation}\label{e9}
g(t,\alpha)=\sum_{l}^{\infty}g_{l}(t)\cos(l\alpha),
\end{equation}

with $l=i\cdot\vert G\vert$  ($l=0$, $\vert G\vert$, $2\vert
G\vert$, {\em etc.}).  Here the angle $\alpha$, describing
nonequivalent directions, is changing between $0\le \alpha \le
\alpha_{G} = \pi/\vert G\vert$. Of course, to use the symmetry
most profitably, a sensible choice is to select the planes
perpendicular to the main axis of the crystal rotation ($\vert
G\vert=4$ or $6$ for cubic and tetragonal or hexagonal structures,
respectively) where $\alpha_{G}$ is minimal, i.e. the number of
equivalent directions is maximal.

Knowing that $l/2$ first coefficients $c_{lk}$ are equal to zero (Eq. (7)),
the equations (8) and (9) give the following dependences between $c_{k}(\alpha)$
\begin{description}
\item[$1^{0}.$] $c_{k}(\alpha) = c_{0k}$, i.e. $\vert G\vert/2$  first coefficients
$c_{k}(\alpha)$  are the same for all projections.
\item[$2^{0}.$]  $2c_{k}(\alpha_{G}/2) = c_{k}(\alpha)+c_{k}(\alpha_{G}-\alpha)$ for
$k \le \vert G\vert$.
\item[$3^{0}$.]  $2\{c_{k}(\alpha_{G}/4)+c_{k}(3\alpha_{G}/4)\}=
c_{k}(\alpha)+c_{k}(\alpha_{G}/2-\alpha)+c_{k}(\alpha_{G}/2+\alpha)+c_{k}(\alpha_{G}-\alpha)$
for $ k \le 2\vert G\vert$.
\end{description}
All these conditions have been proved for both various models of $\rho({\bf p})$
and for $\vert G\vert=2$, $4$ and $6$.

\subsection{Plane projections}

Due to the symmetry of  $\rho({\bf p})$, in the case of  $N=3$,
$g(t,\mbox{\boldmath $\zeta$})\equiv g(t,\Theta,\varphi)$  can be expanded  into
the lattice harmonics $F_{l}(\Theta,\varphi)$ which
form an orthogonal set of linear combinations of the spherical harmonics
$K^{m}_{l}=c(l,m)P^{\vert m\vert}_{l}(\cos\Theta)(e^{im\varphi}+e^{-im\varphi})$.
$c(l,m)$ is the normalization constant and
$P^{\vert m\vert}_{l}(\cos\Theta)$
denotes the associated Legendre polynomial. Angles $(\Theta,\varphi)$ describe the
azimuthal and the polar angles of the $\mbox{\boldmath $\zeta$}$-axis with respect to the
reciprocal lattice system.

In the case of the hexagonal structure $F_{l} = K_{l}^{m}$ with $l=0$ $mod$ $2$ and
$m =0$ $mod$ $6$  ($K_{0}^{0}=1$,  $K_{2}^{0}$, $K_{4}^{0}$, $K_{6}^{0}$, $K_{6}^{6}$, $K_{8}^{0}$,
$K_{8}^{6}$, {\em etc}.), for the tetragonal structure $F_{l}=K_{l}^{m}$
with $l=0$ $mod$ $2$ and $m=0$ $mod$ $4$ and for the cubic structures $F_{l}$ are the
linear combinations of $K_{l}^{m}$, where the first three are equal to [9]:

$F_{0}=1$,

$F_{4}=0.76376261K_{4}^{0}+0.64549722K_{4}^{4}$,

$F_{6}=0.35355338K_{6}^{0}-0.9354134K_{6}^{4}$.

So, for the hexagonal and tetragonal structures a few first lattice
harmonics do not depend on $\varphi$. Moreover, very often, for some paticular
sets of measured spectra, we cannot calculate functions $g_{l}(t)$ (the matrix
 in Eq.~(2) is singular).
 For example:  if for the hexagonal structure we have only three
 projections with $\mbox{\boldmath $\zeta$}$ along the $\Gamma M$, $\Gamma K$ and $\Gamma A$
 symmetry directions (Fig.~1), the angles $(\Theta,\varphi)$ are equal to:  $(\pi/2,\pi/6)$,
 $(\pi/2,0)$ and $(0,0)$, respectively. Thus, the first three lattice harmonics have
 the same values for the $\Gamma M$ and $\Gamma K$ directions and there is no
 a possibility to evaluate functions $g_{l}(t)$. Next, if e.g.
 $(\Theta,\varphi)= (0,0)$, $(\pi/4,0)$, $(\pi/2,0)$,
 $(\pi/4,\pi/6)$, $(\pi/2,\pi/6)$, the set of equations for $g_{l}$  does not
 have a solution in spite of the fact that the first five harmonics
 distinguish here all directions
 $\mbox{\boldmath $\zeta$}$.  Due to these reasons we propose to treat plane projections of
 $\rho({\bf p})$ as line projections of $\tilde\rho_{L}$  and to use the
 consistency and symmetry conditions derived in Sec.~II.A.

 We consider only those spectra $g(t,\mbox{\boldmath $\zeta$})$ for which $\mbox{\boldmath $\zeta$}$ is
 changed on the plane perpendicular to the axis  of the crystal rotation of
 the order $\vert G\vert$. In such a case $\tilde\rho_{L}$ denotes the line
 integral of $\rho({\bf p})$ along lines $L$ parallel to this axis and
 $g(t,\mbox{\boldmath $\zeta$})$ can be described by Eq.(9).  Now all relations obtained for
 the line projections are valid
 with $c_{k}(\alpha)$ being replaced by $c_{k}(\mbox{\boldmath $\zeta$})\equiv c_{k}(\Theta,\varphi)$
 where either $\alpha=\varphi$ and $\Theta=\pi/2$ (for the planes perpendicular to
 the main axis of the crystal rotation)  or $\alpha=\Theta$  and $\varphi= const.$
 (for the planes with $\vert G\vert=2$, perpendicular to the previous ones).

 For example, for three spectra having the hexagonal symmetry and with
 $\mbox{\boldmath $\zeta$}$ along the $\Gamma M$, $\Gamma K$ and $\Gamma A$
 symmetry directions (denoted here by $g(\Gamma M)$, $g(\Gamma K)$
 and $g(\Gamma A)$) we have the following relations: $c_{k}(\Gamma M)=c_{k}(\Gamma K)$ for
 $k=0$, $1$, $2$ (here $\vert G\vert=6$) and $c_{0}(\Gamma A)=c_{0}(\Gamma M)
 =c_{0}(\Gamma K)$ ($\vert G\vert=2$), where  $c_{0}$ denotes the norm of the
spectrum. So, to get more information about $g(\Gamma A)$ we should have at least one
additional spectrum  $g(\Gamma I)$ (for $\Theta=\pi/4$ and e.g. $\varphi=0$)
on the plane with $\vert G\vert=2$ where $2c_{1}(\Gamma I)=
c_{1}(\Gamma A)+c_{1}(\Gamma K)$ (see Fig.~1). This equality is satisfied for $\Theta=\pi/4$
and any $\varphi\in(0,\pi/6)$, i.e. $2c_{1}(\Gamma I'')=c_{1}(\Gamma A)+c_{1}
(\Gamma N)$, where the $\Gamma N$ and  $\Gamma I''$ directions are defined by
the same $\varphi$ with  $\Theta=\pi/2$ and $\Theta=\pi/4$, respectively. Knowing that
$c_{1}(\Gamma N)=c_{1}$ (does not depend on $\varphi$) we obtain that
$c_{1}(\Theta=\pi/4,\varphi)$ has the same value for each spectrum
$g(\mbox{\boldmath $\zeta$})$ with $\mbox{\boldmath $\zeta$}$ described by $\Theta=\pi/4$ and any $\varphi$.
This is derived by treating $g(\Gamma N)$ simultaneously as the line projection
of $\tilde\rho_{\vert 6\vert}$   and $\tilde\rho_{\vert 2 \vert}$ with the symmetry $\vert G\vert=6$
and $\vert G\vert=2$, respectively. Here $\tilde\rho_{\vert 6\vert}$ represents the line
projection of $\rho({\bf p})$ along the main axis of the rotation, while
$\tilde\rho_{\vert 2\vert}$  along any line perpendicular to this axis.
This last dependence is particularly interesting because it gives interdependence
between line projections of different densities, while all consistency conditions
for the line projections are for the same $\tilde\rho$.

All dependencies, shown here on the example of  the hexagonal  structure, are valid
for the cubic and tetragonal structures where $\vert G\vert=6$ is replaced by
$\vert G\vert=4$. However, for  the cubic structures, where three axes of the
fourth order exist, one can get some additional rules.  Because directions
$(\Theta,\varphi)=(0,0)$  and $(\pi/2,0)$ are equivalent, we obtain
that not only $c_{0}$  but also $c_{1}$ must be  the same for all projections.
Some of these results can be also derived from Eq.~(2) because for the
cubic structures all lattice harmonics (except for $F_{0}$) depend on $(\Theta,\varphi)$.
 Taking as an example directions $[00h]$, $[kl0]$ and $[klh]$ we obtain the following
 dependences between $c_{i}$:

$a_{1}c_{i}([00h])+a_{2}c_{i}([kl0])+a_{3}c_{i}([klh])=0$      for   $i=0,1$

$b_{1}c_{i}([00h])+b_{2}c_{i}([kl0])+b_{3}c_{i}([klh])=0$     for   $i=0,1,2$
    where $\sum b_{i}=\sum a_{i}=0$.

The above equalities arise from the conditions of vanishing all polynomials
$W_{i}$ with $i=0,1$ and $i=0,1,2$ in the expansion of $g_{4}(t)$ and $g_{6}(t)$,
respectively. The coefficient $c_{1}$ satisfies both equations when $c_{1}([00h])=
c_{1}([kl0])=c_{1}([klh])$
what is in agreement with the previous result. Knowing additionally that
$c_{1}([kl0])$ does not depend on the direction $[kl0]$, we can conclude
that for the cubic structure the value of $c_{1}([klh]$) does not
depend on the direction (denoted here by $[klh]$).

As before, all relations were proved for both model and experimental profiles.
Some examples are presented in the next section.

\section{Application}

First, disposing twenty five $2D$ $ACAR$ experimental spectra [10] for the four metals of the hcp structure
with  $\mbox{\boldmath $\zeta$}$  on the plane $\vert G\vert=6$, we obtained
that the
conditions $1^{0}$  and $2^{0}$ are satisfied with a very high accuracy,
lower than $0.5\%$  ($c_{1}$), $1.5\%$  ($c_{2}$), $1.5\%$  ($c_{3}$),
$2.5\%$  ($c_{4}$) and  $5.5\%$  ($c_{5}$).
Because the first $ck{m}(\mbox{\boldmath $\zeta$})$ have the highest values,
they are determined the best (the influence of the statistical noise is the
lowest).
This very small inconsistency implies that spectra [10] (with the average number of counts at peak about
60000) were not only measured  but also corrected with a very high precision.

In the figure 2 we show results for the theoretical Compton profiles
for Cr [11] with  $\mbox{\boldmath $\zeta$}$ along $[100]$, $[110]$ and $[111]$ directions.
In this case (cubic structure) the first two coefficients $c_{k}(\mbox{\boldmath $\zeta$})$ should have
the same value for each spectrum, while the third coefficient should satisfy
the relation: $c[100]-c[110]=3\{c[110]-c[111]\}$. After normalizing spectra to the
same area ($c_{0}=1$) we obtained that a distortion of $c_{1}$ from its average
value is changed from $0.1\%$ up to $0.4\%$ and the inconsistency of $c_{2}$ is of
order $1.5\%$. Thus, the behaviour of $c_{k}(\mbox{\boldmath $\zeta$})$
is the same as for previously studied experimental data. It is connected with the
fact that in order to create such spectra it would be necessary to calculate $\rho({\bf p})$ in the
whole momentum space ${\bf p}$, what, of course, cannot be done.  Here
the accuracy of calculating Compton profiles was comparable with the statistical
noise of 2D ACAR data [10]. Of course, in order to get the similar inconsistency
of the Compton profiles
(where a contribution of core densities is much higher than in a positron annihilation
experiment) the experimental statistic must be also much higher.

In order to examine how these conditions react to an improper shape of
$g(t,\mbox{\boldmath $\zeta$})$, we changed somewhat the shape of one spectrum
( marked in figure 2 by squares). Here we would like to point out
that this incorrect spectrum does not differ from other spectra too much
(the differences between $g([100])$ and $g([111])$ are higher). After
normalizing it to the same area ($c_{0}=1$) we obtained
that $c_{1}$ changed its value from -0.7 to -0.74 . So, here
we observe a much higher inconsistency ($5\%$)  than for
experimental data [10] ($0.5\%$).

\section{Conclusions}

All spectra received from experiments are contaminated by statistical noise
and thus they are not consistent.  Depending on the number of measured
projections, some part of  the inconsistent noise is eliminated via the
consistency conditions  during the reconstruction of $\rho({\bf p})$.
However,  we propose to check (before the reconstruction) if spectra were
measured and corrected properly (the inconsistent part of the data cannot
be too large). For that purpose we can use the interdependences, derived by
combining the consistency
and symmetry conditions, obtained here for the line and plane projections.

The knowledge of the interdependences rules can be also profitable for an
{\em improvement} of data for which densities are not reconstructed (some part of
the inconsistent noise can be eliminated). Moreover, it can be applied to
verify these techniques which are used for correcting
Compton scattering spectra (they are not univocal and
can be individual for each spectrum [12]). For that we propose to measure
the high resolution {\em CP}, denoted usually by $J(p_{z})$, for hexagonal metals
with $p_{z}\equiv\mbox{\boldmath $\zeta$}$ changed on the plane $\vert G\vert=6$ ($\Theta=\pi/2$). Having $p_{z}$ along $\Gamma K$ ($\varphi=0$),
$\Gamma M$ ($\varphi=\pi/6$) and $\Gamma N$ ($\varphi=\pi/12$), we can check if
$c_{i}(\Gamma K)=c_{i}(\Gamma M)=c_{i}(\Gamma N)$ for $i=0, 1, 2$ and if
$c_{i}(\Gamma K)+c_{i}(\Gamma M)=2c_{i}(\Gamma N)$ for $i=3, 4, 5$. Of course, the
best choice is to study rare-earth metals where the anisotropy of $\rho({\bf p})$
is so high that spectra $J(p_{z})$ should be essentionally different.

\section*{Acknowledgements}
I am very grateful to Professor R. M. Lewitt for helpful discussions, Professor R. N. West for
making available his experimental 2D ACAR data and to the State Committee for
Scientific research (Republic  of Poland, Grant No 2 P03B 083 16) for financial
support.

\begin{figure}[htb]
\scalebox{0.6}{\includegraphics{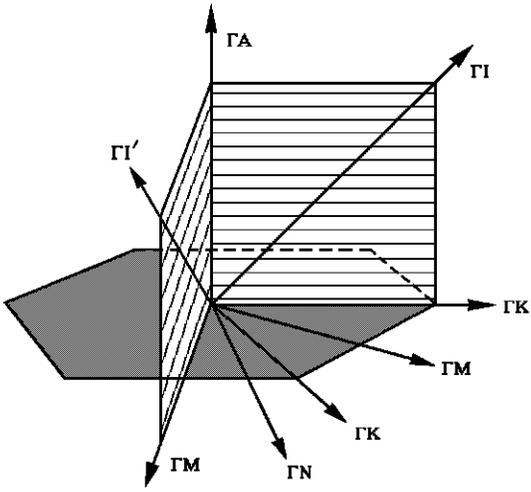}}
\caption{Planes and directions for the hexagonal structure having symmetry of the
sixth and the second order, marked by dots and lines, respectively.}
\label{fig1}
\end{figure}

\vskip 1.0truecm

\begin{figure}[htb]
\scalebox{0.33}{\includegraphics{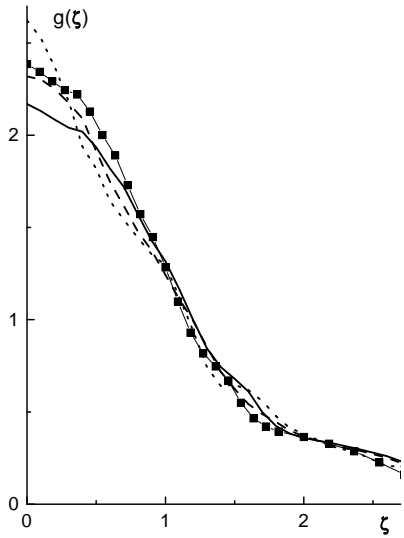}}
\caption{Theoretical plane projections for Cr
[11] having cubic symmetry with  ${\bf \zeta}$  along [100], [110]
and [111] directions marked by full, broken and dotted lines,
respectivelly. {\em Incorrect} projection [110] is marked by the
full squares.} \label{fig2}
\end{figure}

\end{document}